\documentclass[review]{elsarticle}

\usepackage{hyperref, boolexpr, amsmath, float}

\journal{Journal of \LaTeX\ Templates}

\begin{document}

\begin{frontmatter}

\title{The alternative bases of Boolean functions as a means of improving the structure of digital blocks}
\tnotetext[mytitlenote]{Contact information.} 

\author{Sergii Kushch}
\ead{kushch@yaros.co, skushch@fbk.eu}

\address{Bruno Kessler Foundation, via Sommarive, 18, Povo, 38123,TN, Italy}

\begin{abstract}
This paper analyzes three forms of representation of Boolean functions, such as Classical, Algebraic and Reed-Muller. The concept of intersection and subsets of representation forms have been introduced, moreover suitable criteria for creating these subsets have been established. Later, these subsets have been quantitatively compared by the number of parameters, in order to assess the effectiveness of using each of the forms of representations proposed in the work. Definitions of the specific weight of subsets of priority forms of the representation of Boolean functions showed that the classical form is the least optimal, in comparison with the parameters of other forms Also, it has been shown that the use of alternative forms of representation of Boolean functions, in some cases, allows to reduce twice the number of incoming PLA buses. Estimating the average loss from the exclusive use of the Classical Form Representation also shows that the use of alternatives yields significant benefits in some parameters, this can be used to optimize devices in the logic design process and reduce the chip area, what also contributes to reductions in the cost of such devices.\\
\end{abstract}

\begin{keyword}\texttt{Boolean Functions \sep Logical Design \sep Representation Form of Boolean Functions \sep Efficiency of Boolean Function Bases}
\end{keyword}

\end{frontmatter}

\section{Introduction.}
It is already a known fact that there are several bases with which any Boolean function can be represented. The present paper is a study of practical consequences of the isomorphism of Boolean functions (BF).  In addition to the classical representation of BF in the form of disjunctive and conjunctive normal forms (DNF and CNF), alternative implementations are also possible.  Moreover, isomorphism, indicates that all the variety of problems and the methods of solving them, are the same for both classical forms of representation and alternative forms. The paper shows that the PLAs, implementing currently BF as DNF use two halfmatrix PLA1 and PLA2 \cite {Roth14, Zosimo15, Balabanian16, Rajaraman17}, respectively, with conjunctions of input arguments and disjunctions of the obtained conjunctions, provide a minimum area of  PLA in less than a third of cases. In other cases, it is expedient to use alternative forms of representation to minimize the PLA area. Necessary conjunctions are realized in PLA-1, and these conjunctions are summed in PLA-2, on summation elements (ESС), which in the classical case are OR elements. There are two large sections in the general theory of automata: the abstract and structural theory of digital automata (DA). In the abstract theory of DA, transitions of the DA are studied under the influence of input signals but without the structure of automata. An important special case of DA are automates with only one internal state, which are called automates without memory or combinational circuits (CC). Combinational circuits are the content core of any DA. A system of logical elements is said to be functionally complete if there is a general method allowing to construct any combination scheme and, in particular, any logical (Boolean) function, only from the elements of this system. At present, the list of functionally complete systems (among logical elements) is exhaustive. From this list, the most well-known systems consist of the following logical functions (LF):\\ 
- conjunction+negation (AND-NOT);\\
- disjunction+negation; (OR NOT);\\
- conjunction+disjunction+negation (AND-OR-NOT);\\
- conjunction+sum mod2+1 (AND-MOD2-1).\\
The BF system (logical basis) AND-OR-NOT Is referred to as the classical form of representation. The system BF - AND-MOD2-1 is known in \cite{Zhegalkin10, Harking11, Swamy12, Saluja13} as the Zhegalkin algebra, also the Reed-Muller Can be referred to as the Reed-Muller form of representation (RMFR), since Zhegalkin polynomials are a particular case of set of Reed-Muller's polynomials. The Algebraic form of representation (AFR) BF \cite{Golomb1} is a result of F-transformation of BF into equivalent piecewise constant functions. Apart from these forms there are also others e.g. Cognate RF \cite{Kochkarev21}, orthogonal AF \cite{Kochkarev22} et al., which we call Alternative Forms.
In this paper, by comparing CFR, AFR, and RMFR, it will be shown that the use of different forms of representation of Boolean functions can yield significant benefits in the logical design when compared with the now widely used CFR BF.\\
Examples of the possible use of such forms are shown in works \cite{Kochkarev19, Kochkarev20}, where  possible variants from constructing adders mod2 using RMFR, are presented. In this case, the signal x at the input of these elements was in only one form - direct or inverted. This is different from the circuits which use the KFR, where the signal must be fed to the inputs of the circuit in both the forward and inverse form.\\
The paper is organized as follows. In Section 2 we formulate the problem and make an overview of the various forms of representation of Boolean functions. The research problem is formulated and ways to solve it are considered. Two alternative forms are evaluated: the Algebraic form of representation (AFR), in which BFs are represented as polynomials with algebraic summation of special S-functions and the polynomial representation form with summation of S-functions by mod2, which are known as Zhegalkin and Reed-Muller polynomials (RMFR) and the difference between them and the CFR BF.\\
The results obtained in the study are outlined in Section 3. A numerical comparison of the effectiveness of different forms is carried out according to the main criteria. A comparison of the power of the so-called priority subsets for each form of representation was conducted in order find out about the cardinality of the cardinality of BF subsets for which the Classical or Alternative implementation of BF is appropriate and, moreover, to know what are the possible reductions of the PLA area when applying this method. It has also been  quite rigorously proved that the powers of subsets of BF, for which it is expedient to use Alternative forms of representation, have the same order as the power of the subset for which CFR is appropriate. Consequently, the current system of implementing BF in CFR provides an optimal solution in less than 50\% of cases. Finally, we present the conclusions obtained from our research and discuss the possibilities for future work in Section 4.\\

\section{PROBLEM FORMULATION}
In this section, we briefly consider the essence of the proposed alternative forms of representation of Boolean functions and the differences between them. It will also be shown how the sets and subsets of these representation forms of BFs interact. In addition, we will formulate the task and the grounds for choosing the most appropriate FR from the given parameters.

\paragraph{\textbf{2.1 Algebraic form of representation}}
The ability to characterize Boolean functions with the help of a certain set of real numbers was first observed in \cite{Golomb1,Givant18}. Later, a representation of Boolean functions as finite sums of Walsh functions was used in \cite{Dertouzos2} for the tasks of logical network synthesis  which use threshold elements. In \cite{Karpovskii3}, a representation of a system of logic functions of arbitrary dimension with the help of the Vilenkin-Chrestenson functions was introduced. The transition from the BF system with n arguments to the piecewise constant function $Ф(x)$ of one continuous argument which varies over the interval $[0, 2^n-1)$ in \cite{Kochkarev4} is called the F-transformation. Ibid considered a class of so-called " Ortofunctional Transformations" (OF-transformation), in which F-transformation is a special case. In this class of OF transformations, two large subclasses are distinguished:\\
-	transformations with a canonical metric in which the closeness between the original and the OF-image is considered in the sense of a minimum of the mean-squared error, i.e. in the metric of space - L2. The set of transformations refers to the indicated subclass, for example, the transformations of Fourier, Walsh, Haar, Chebyshev, Hermite, Laguerre, Legendre and, in particular, the considered F-transformation.\\
-	transformations with a non-canonical metric in which the closeness between the original and the OF-image is taken in a form specially specified for each particular transformation. The K- and P-transformations, introduced in \cite{Kochkarev4}, belong to this subclass.\\
It should be noted that the use of the canonical F-transformation of BF is not always convenient, because it does not always provide a minimal solution for a circuit. In this regard, the task of reducing the circuit complexity in implementing the BF becomes relevant. The possibility of minimization consists of the F-transformation of BF with non-canonical metric. For such a case, we associate with the original $f(x_1, x_2,\ ...\ , x_n)$ the grid function $Z(x)$ defined at the points $x=0,1,2, ... ,2^n-1$, as for the canonical F -transformation, the numbers of which, for all possible sets of arguments, are determined by the formula: $x=\sum \limits_{s=1}^{n-1} x_s 2^{n-s}$. The values of $Z$ are equal $f(x_1, x_2, ..., x_n)$ to the corresponding sets of arguments. Furthermore, for the resulting net function $Z(x)$, given at $2^n$ points and assuming the values 0 and 1, we assign a piecewise constant function $F(x)$ defined on the interval $[0,2^n)$. It has $2^n$ unit unit intervals of constancy, and on any, for example, a unit interval of constancy, it is needed the fulfillment of the condition:
$F(\delta)>0,5\ if\ Z(\delta=1)$;
$F(\delta)<0,5\ if\ Z(\delta=0)$.\\
This condition is the main difference between the non-canonical F-transformation and the canonical one. If earlier the coincidence of $Ф(x)$ and $Z(x)$ was necessary at the points  $0,1,2,\ ...\ 2^n-1$, now only the nonnegativity of $F(x)-0.5$ is necessary for $Z(x)=1$ and the nonpositiveness of $F(x)-0.5$ for $Z(x)=0$.\\
The orthofunctional F-transformation of truth tables (TT) that defines the combinational circuit of $K$-valued logic to the corresponding functional series, served as a basis for considering the isomorphism BF between CFR and other (alternative) bases. Further, Boolean function are named identical (equivalent) only if they have the same truth tables.  In this regard, BF, recorded in different logical bases, but having the same TT, will be considered identical. \\
Two sets $R_1$ and $R_2$ re isomorphic with respect to some operations in the indicated sets if for some $R_1$ and  $R_2$ elements there is a one-to-one correspondence r, of the following form:
\begin{math} \left.\begin{matrix}
r(f_{i})=\Phi_{i} \\
r^{-1}(\Phi_i)=f_i
\end{matrix}\right\}f_i \in{R_1}, \Phi_i \in{R_2} \end{math}.
Also the following correspondence: $r(f_i\Delta f_k)=\Phi_i \diamondsuit{\Phi_k}$, $r^{-1}(\Phi_i \diamondsuit \Phi_k)= f_i\Delta f_k$. Here, the symbols $\Delta$ and $\diamondsuit$ designate certain operations specified in the sets $R_1$ and $R_2$, respectively. Let's look at the functions written in the logical basis and the functions corresponding to them, written using an orthogonal basis for the general case of $k \ge 2$, as the sets $R_1$ and $R_2$. We will consider the set of functions of $k$-valued logic ($K \ge 2$) for $n$ arguments ($n \ge 1$), as the set $R_1$. As operations defined on the set $R_1$, choose, for example, the Rosser-Tuckett system, which was developed \cite{Pospelov5} The specified system contains the following elements: constants - $f_j=j, 0 \le j \le k-1$; characteristic functions of one argument:\\ 
$
\Psi_j(x)=
\left\{\begin{matrix}
k-1, if\ x=j \\
0\ if\ \ x \not= j
\end{matrix}\right.
$ 
and the following logical operations: disjunction - $f_i \lor f_j =max(f_i, f_j)$; conjunction - $f_i \land f_j =min(f_i, f_j)$. As the set $R_2$, we consider the set of functions $Ф(x)$ of one argument, given on the interval $[0,k^n)$ which has $k^n$ identical unit intervals of constancy. The function $Ф(x)$ takes only one value from the set  ${0,1,2,…,k-1}$, at each of these intervals. This set - is finite, it consists of $k^{k^n}$ elements and each of its elements is an F-image of a TT which has $k^n$ rows and $n+1$ columns. The following operations are defined on the indicated set: graphical disjunction - 
$\Phi_i(x) \oplus \Phi_k(x) =0.5[\Phi_i(x)+\Phi_k(x) + \mid {\Phi_i(x)-\Phi_k(x)} \mid]$; graphical conjunction - $\Phi_i(x) \otimes \Phi_k(x) =0.5[\Phi_i(x)+\Phi_k(x) - \mid {\Phi_i(x)-\Phi_k(x)} \mid]$. The operations $\bigotimes$ and $\bigoplus$ are expressed in terms of known algebraic operations of summation and multiplication. The isomorphism of the indicated sets $R_1$ and $R_2$ with respect to the operations of disjunction and conjunction is proved in \cite{Pospelov5}. Note, that in the case of binary logic:

\begin{equation}\label{formula_1}
if \; f(x_1, x_2, ... , x_n) \approx \Phi(x) \; to \; 
F(x_1, x_2, ... , x_n) \approx {1-\Phi(x)} = {\Phi_n(x)}
\end{equation}

In \cite{Kochkarev6}, various systems of basis functions were introduced, for the convenience of an analytic representation of F-images. Considering, in particular, a convenient system, which is hereinafter referred to as an S-basis. The system of S-functions of the $n^{th}$ order consists of $2^n$ functions and for a different number \boldmath$n$ it can be constructed recurrently.  

\begin{equation}\label{formula_2}
\left\{\begin{matrix}
 S_{00}=1 \\
 S_{jn}=S_{j,n-1}, \; if \; 0 \le j \le 2^{n-z}-1\\
 S_{jn}=X_{n}S_{j-2^{n-1},n-1} \; if \; 2^{n-1} \le j \le -1 \\
\end{matrix}\right.
\end{equation}

The use of the S-basis (\ref{formula_2}) corresponds to the BF representation, or more precisely, it's F-image, in AFR, i.e. in the form:

$f(x_1, ... , x_n) \sim C_0 + C_1X_1+ ... + C_{123...n}X_1X_2...X_n = \\ 
=\Phi(X_1, X_2, ... X_n)=\sum \limits_{j=0}^{j^n-1}{C_jS_j(x)}$,

where $x_i$ represents the arguments of the BF; $X_i$ - corresponding to these arguments in CFR. Every BF of n arguments can be represented as a linear combination of S functions (S-series) using this basis. The coefficients of this series can be regarded as the coordinates of the vector in the $2^n$ -dimensional space of the CFR. Thus, any BF of n-variables can be uniquely represented by the vector $\bar{C_s}$ in the S basis. This basis is not the only possible one. A Q-basis of a system of q-functions, defined in the same $2^n$-dimensional CFR space, is introduced in \cite{Kochkarev6}. Note, that the coordinates of the BF in the Q-basis are actually the BF values on all $2^n$ sets of arguments in the TT. It is possible to write every BF as a set with no more than $2^{n-1}$ using the methods described above. The set consists of $2^n$ terms, for other bases, in the general case. As a comparison, we note that for the expansion of an arbitrary CFR $\phi(x)$ on the Walsh basis, it is required to perform $n \cdot 2^n$ addition-subtraction operations, and for the Haar-basis $2^{2n-1}$ operations.\\
Proceeding from the foregoing, the transition from CFR to AFR consists in the transformation of the coordinates of the BF from the S-basis and the Q-basis.

\paragraph{\textbf{2.2 Reed-Muller form of representation}}
Another algebraic system, alternative to classical, is a system based on the operation of summation over mod2. Equivalence relations between these two algebraic systems can be obtained on the basis of theorems and identities of Boolean algebra. It should be noted that both these algebraic systems have a common multiplicative operator AND (conjunction). The additive operator is implemented in Boolean algebra by the OR function (OR, disjunction), and in the Zhegalkin algebra by the exclusive OR function (XOR, sum mod2). The basic relation between algebraic systems is expressed by the identity:
$x \vee y= x \oplus y \oplus x \cdot y$.
The polynomial representation for the $i^th$ BF, in the general case, has the form:
$f(x_1, x_2, ... , x_n)=F_{i1} \oplus F_{i2} \oplus ...  \oplus F_{i_j}= \bigoplus \limits_{i{_j}}{F_{i_j}}$. Here: $i_j \in T1$; $T_1$ is the set of numbers of sets on which the function becomes 1; $F_{i_j}$ is is the complete conjunction, i.e. of the constituent 1 on the j-th set.\\
Distinguish the following cases of polynomial representation:\\
- polynomial perfect normal form (PPNF) \cite{Pospelov5}, obtained from the PDNF / perfect disjunctive normal form / by replacing the disjunction by adding mod2:\\
$f(x_1, ... , x_n)=\bigoplus \limits_1{x_1^{\alpha_1}x_2^{\alpha_2}, ... , x_n^{\alpha_n}}$. Here the symbol $\bigoplus \limits_1$ means that the sum over (mod2) is taken only by such sets $<\alpha_1, \alpha_2, \alpha_3>$, on which the BF is equal to one.\\  
-  the canonical Zhegalkin polynomial which is obtained from PPNF by the elimination of inversion $\bar{x}=x\oplus1$ and reduction of similar elements. In the algebra of Zhegalkin, the role of perfect forms of Boolean algebra is played by canonical polynomials. A canonical polynomial is a finite sum of pairwise distinct product of variables, such that in one and the same product any variable exists not more than once. In this case, a product consisted only a one cofactor (individual variables), and a product consisting of an empty set of factors (constant 1) also belongs to the set of products. To the number of elementary polynomials for the completeness of the system, we must also include the constant 0, considering it as a sum mod2 of an empty set of terms.\\
- canonical polarized polynomials \cite{Zakrevskii7}, which are a generalization of Zhegalkin polynomials, where the recording and operations on BF are carried out in an algebraic system mod2 taking into account the so-called polarization vector. The representation of BF in the form of polynomials of this type is called the Reed-Muller polynomial, here Reed-Muller form of representation (RMFP) \cite{Sasao8}.
The Reed-Muller polynomials can be written in general form:

\begin{equation}\label{formula_3}
f(x_1, x_2, ..., x_n)=a_0g_0\oplus a_1g_1 \oplus ... \oplus a_ng_n
\end{equation}

Where $g_0,g_1,…,g_k$ conjunctions of input arguments or their negations;\\
$a_0,a_1,…,a_k$ are the coefficients of the polynomial taking the value 0 or 1.
Some or all of the arguments of the BF in (\ref{formula_3}) can be inverted (in terms of \cite{Zakrevskii7} they are polarized), and each argument can be in only one (direct or inverse) form. The polarization vector is an ordered set of zeros and ones, in which the inverted input arguments correspond to ones.  Expression (\ref{formula_3}) in expanded form:

\begin{equation}\label{formula_4}
\begin{matrix}
f(x_1, x_2, ..., x_n)=a_0\oplus a_1 \tilde{x}_1 \oplus ... \oplus a_n \tilde{x}_n \oplus a_{12} \tilde{x}_1x_2 \oplus \\
\oplus a_{x,n-1} \tilde{x}_n \tilde{x}_{n-1} \oplus ... \oplus a_{12...n} \tilde{x}_1 \tilde{x}_2... \tilde{x}_n
\end{matrix}
\end{equation}

All Reed-Muller polynomials can be classified taking into account the character of the polarization vector as follows:\\
- Reed-Muller polynomials of positive polarity, in which the variables enter only in the direct form (the polarization vector contains only zeros). This form of representation has been described above as Zhegalkin representation form.\\
- Reed-Muller polynomials of negative polarity, i.e. when the variables enter only in inverse form (the polarization vector contains only units).\\
- Reed-Muller polynomials of fixed polarity, defined by a fixed polarization vector, when each variable can enter only in one form (either in direct or in inverse form).\\
It is obvious that polynomials of positive and negative polarity are particular cases of polynomials of fixed polarity. The uniqueness of polarized polynomials is provided by specifying the polarization vector $V_k \sim X=(\tilde{x}_1 \tilde{x}_2... \tilde{x}_n),\ k=0 \div 2^n-1$ which can be given by a decimal integer or an n-bit binary vector. The value of the component $\tilde{x}_i$ of the polarization vector determined by the value of the argument $x_i$: 0 - direct, 1 - inverse. \\
For the given fully defined BF and a given VP $V_k$, here exists a unique Reed-Muller polynomial. The number of possible polarized vectors $V_k$ for a function of n variables is equal to $2^n$, which determines the total number of possible Reed-Muller polynomials of fixed polarity. Then for the whole set of BFs of $n$ variables the total number of Reed-Muller polynomials is equal to  $2^n \times 2^{2^n}$. It should be noted that among this set for each BF there are one or several forms that provide the minimum form of representation (according to the chosen criterion). In addition to the above classification, among the set of Reed-Muller polynomials one can distinguish classes of polynomials with allowance for the length of the conjunction [8], that is, the number of variables which are included in these conjunctions. The indicated classes (subsets) of the Reed-Muller polynomials have the form: $E_0^n=1$; $E_1^n= \sum \limits_{i=1}^n {\oplus x_i}$; $E_2^n= \sum \limits_{1 \le i < j \le n} {\oplus x_ix_j}$; $E_k^n= \sum \limits_{1 \le i < j <...<m \le n} {\oplus x_ix_j, ... x_m}$; $E_n^n= x_1 \cdot x_2 \cdot x_3 \cdot ... \cdot x_n$. For every BF of n arguments, the total number of different classes is $n$, and the zero class $E_0^n$ consists of a single polynomial equal to the value of unity, the class $E_n^n$ onsists of the unique conjunction of all arguments. With this in mind, we can estimate the power of each class, i.e. the limiting number of summands, as $E_0^n=E_n^n=1$, $E_1^n=n$, $E_k^n=E_k^{n-1}+E_{k-1}^{n-1}$. Let us further consider the possible intersections and interactions of the sets of these FRBFs with each other.

\paragraph{\textbf{2.3 The intersections and the subsets of the FRBFs}}
Note, as an important fact, that BFs, which are implemented in the simplest way in one FR - in other forms, as a rule, require the most complex realizations.\\
In \cite{Kochkarev9} it was suggested to split the complete set of BFs from n arguments (we denote it as L(n)) by several subsets. Let us consider the possible subsets of intersections of CFR, AFR, and RMFR:
\begin{itemize}
\item C is a subset of BF, for which the CFR is most suitable;
\item A - subset of BF, for which the AFR is most suitable;
\item RM is a subset of BF, for which the RMFR is most suitable.
\end{itemize}
In addition, we introduce intermediate subsets:
\begin{itemize}
\item CA is a subset of the BF, for which both CFR and AFR are equally useful;
\item CR - a subset of BF, for which both CFR and RMFR are equally advisable;
\item AR - a subset of BF, for which both AFR and RMFR are equally useful;
\item CAR - a subset of the BF, for which any form of representation is equally appropriate. 
\end{itemize}

\begin{figure}[H]
\centering{\includegraphics[width=50mm]{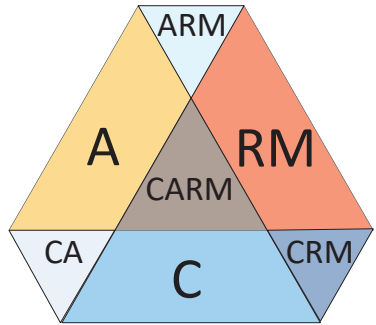}}
\caption{\textbf{The Venn diagram for the set of BFs L(n) of n-arguments.}\label{figure1}}
\end{figure}

Naturally, the question arises about the cardinality of all the above subsets of priorities (SP). This question is quite important from both the theoretical and practical point of view.\\
In fact, if the subset of CAR, for example, is predominate on L(n), it is inappropriate to use alternative AFs, since circuitry for them can be somewhat more complicated than for CFR. In other words, the introduction of alternative FRs is expedient if and only if the cardinality of the A- and R-subsets are sufficiently large. Obviously, the power of the SP depends on the power L(n), i.e. of the value of n.\\
We note that not only the absolute values of the power of various SPs depend essentially on the value of n, but also their relative specific weight too. In fact, for $n=1$, the entire set L(1) consists of a single SP CAR. For $n=2$, the set L(2) consists of  SPs of the type CAR, CA, and R.\\
The situation presented above can be illustrated with the help of a Venn diagram (Fig.\ref{figure1}), where participate SPs - C, A, RM, as well as intermediate PPs - CA, CR, AR, CA are represented. It is important to note that with the increase in the number of arguments, the process of differentiating L(n) on the SP becomes increasingly clearer (hereinafter it will be evaluated) quantitatively.

\paragraph{\textbf{2.4 Selection the representation form of BF}}
Let us analyze the situation that arises from the existence of classical (CFR) and alternative (AFR, RMFR) representation forms of Boolean functions. All forms are isomorphic and are the basis for the so-called "concept of Optimal Representation Form". Which consists in choosing the most appropriate RF for a particular BF. It is assumed that the goal of rational choice of the form of representation of the BF is to reduce the area of the PLA, which, as is known, consists of two submatrices. In the submatrix PLM1, conjunctions are formed that are necessary for any form of the representation of the BF, and in submatrix PLM2, the formed conjunctions are summed, logically (in the CFR), algebraically (in AFR) or mod2 (in the RMFR), depending on the chosen FR. In this paper, we will consider the minimization of the area of the submatrix PLM1, as the main submatrix, which affects the area of the PLA. The rational circuitry PLM2 which uses all three FR BFs, is an object of additional research and is not considered in this work. It should be clearly understood that a decrease in the area of PLM1 using alternative FRs will be accompanied by a slight increase in the area of the PLA2 and the issue of the use of a particular FR in a particular case should be addressed in the light of this circumstance. For an implementation of the BF not only on the PLA, it is necessary to take into account a set of criteria that must meet the practical requirements of the design. Naturally, before comparing the different FRs of a particular BF with each other, the BF must be minimized in all forms. Only after this the comparison will be correct.
When deciding the effectiveness of various FRs, the cardinality of the subsets of the priority of a particular FR, and the choice of the optimal FR, the following quality criteria will be used:

\begin{itemize}
\item $S_{ad}$ - the number of summands in the BF record that determines the number of inputs of the submatrix PLM2;
\item $S_{SH}$ - the number of summands in the BF record representing the conjunction of input arguments, which determines the number of lines in PLM1 with sets of active elements;
\item $S_L$ - the number of letters in the BF record, which is a classic criterion for minimizing BF;
\item $S_s$ - overall area of PLM1, which is defined as:
    \begin{itemize}
    \item $S_s=2nS_{ad}$ (for CFR);
    \item $S_s=nS_{ad}$ (for AFR and RMFR), where n is the number of input arguments of PLM1.
    \end{itemize}
\item $S_{ac}$ – an area of active elements of PLM1, defined as:
    \begin{itemize}
    \item $S_ac{ac}=2nS_{SH}$ (for KFR);
    \item $S_{ac}=nS_{SH}$ (for AFR and RMFR).
    \end{itemize}
\end{itemize}
Note that from the point of view of the implementation of a specific BF, the indicated values of the criteria can be considered as the criterions of the complexity of the implementation of the BF, which, naturally, depend on the chosen FR. The presence of CFR, AFR, and RMFR allows us to set and objectively solve the following problems: comparison (quantitative) of the effectiveness of different FR BF; the determination of the specific gravity of various SPs in L(n); Estimation of losses (average) from exclusive use of CFRs.

\paragraph{\textbf{2.5 Mathematical model}}
The presence of alternatives in solving any problem (from the fundamental scientific to the current household) inevitably raises the question of their comparison, and the comparison requires some quantitative evaluation of the alternatives. Taking into account the fact that all the criterions S represent integers, as well as the piecewise-constant nature of the functions Ni(S), the relative efficiency index (REI) of the $i^{th}$ FR can be written in the following equivalent form:

\begin{equation}\label{formula_5}
\eta{_i}=\frac {\sum \limits_{j=0}^{S_{mm}}N_{ij}(S)}{N_{max}S_{mm}}
\end{equation}

Where: $N_{ij}$ - is the number of BFs realized on the PLA, given the value of the specified criterion; $N_{max}$ - the total number of BFs of a given number of arguments n; $S_{mm}$ is the maximum value of the selected criterion for all FRs, which ensures the realization of all BFs.\\
If (\ref{formula_5}) is rewritten in the form (\ref{formula_6}), then REI value has a clear statistical meaning.

\begin{equation}\label{formula_6}
\eta{_i}=\frac 1S_{max} \sum \limits_{j=0}^{S_{max}} {\frac {N_{ij}} {N_{max}}} = 
\frac 1S_{max} \sum \limits_{j=0}^{S_{max}} {p_{ij}}
\end{equation}
\\
It is clear from (\ref{formula_6}) that $p_{ji}$ is the probability of BF realization in the $i^{th}$ FR at the value of the chosen criterion $S \le j$, and the whole value of the REI $\eta_i$  is the average value of probability of the realization of the BF in the $i^{th}$ FR by the selected criterion S.\\
To quantify the losses from exclusive use in modern microcircuits only the classical FR, the sums of the main indicators of the complexity of realization over the complete sets L(3), L(4) and L(5) are calculated:

\begin{equation}\label{formula_7}
Q_{ad}=\sum \limits_{i=0}^{2^{2^n}-1} {S_{ad}^{(i)}}
\end{equation}

\begin{equation}\label{formula_8}
S_s=\sum \limits_{i=0}^{2^{2^n}-1} {S_s^{(i)}}
\end{equation}

\section{MAIN RESULTS}
\paragraph{\textbf{3.1 Quantitative comparison of the effectiveness of BF representation forms.}}
 The Information about the comparative efficiency of the use of different FRs for all criteria noted above was obtained by analyzing BF by the classical and the alternative minimal forms for n=3, 4, 5. The results for the (OFR) (\ref{formula_5}), are also shown. This form of representation corresponds to the PLM, in which the second semimatrix PLM2 contains elements OR, XOR or comparators. This depends on the subset that contains the implemented BF, by the given criterion.\\
The calculations according to (\ref{formula_5}) (\ref{formula_6}) give the parameters of the REI for all the FR BF for n=3, by all selected criteria are given in Table. \ref{table1}

\begin{table}[h!]
\caption{\textbf{Relative performance efficiency of all representation forms for the complete set of BFs, n=3}\label{table1}}
\begin{center}
\begin{tabular}{c|c|c|c|c|c}
& \multicolumn{5}{c}{REI for the quality criterions} \\
\cline{2-6}
\raisebox{1.5ex}[0cm][0cm]{Representation form}
& Sad & Ssn & SL & Ss & Sac \\
\hline
Classical (CFR) & 0.74 & 0.61 & 0.63 & 0.53 & 0.47\\
\hline
Algebraic (AFR)	& 0.58	& 0.8	& 0.61	& 0.63	& 0.89\\
\hline
Reed-Muller (RMFR)	& 0.7	& 0.75	& 0.66	& 0.76	& 0.86\\
\hline
Optimized (OFR)	& 0.76	& 0.86	& 0.71	& 0.82	& 0.92\\
\end{tabular}
\end{center}
\end{table}

From Table \ref{table1}, it follows that the generally accepted CFR only by the Sad criterion can provide an effective implementation, as compared to alternative FRs.\\
A similar analysis of the functions for n=4 in the classical and alternative minimal forms of BFs has been conducted and its results are provided in Table. \ref{table2}.

\begin{table}[h!]
\caption{\textbf{Relative performance efficiency of all representation forms for the complete set of BFs, n=4}\label{table2}}
\begin{center}
\begin{tabular}{c|c|c|c|c|c}
& \multicolumn{5}{c}{REI for the quality criterions} \\
\cline{2-6}
\raisebox{1.5ex}[0cm][0cm]{Representation form}
& Sad & Ssn & SL & Ss & Sac \\
\hline
Classical (CFR) & 0,742 &	0,575	& 0,645 &	0,514 &	0,517\\
\hline
Algebraic (AFR)	& 0,644	& 0,611	& 0,664	& 0,665	& 0,784\\
\hline
Reed-Muller (RMFR)	& 0,656 &	0,591 &	0,642 &	0,676 &	0,770\\
\hline
Optimized (OFR)	& 0,747	& 0,667 &	0,711 &	0,701 &	0,816\\
\end{tabular}
\end{center}
\end{table}

It can be seen that the change in the number of arguments in BF does not change the situation in essence - from the comparison in Tables \ref{table1} and \ref{table2}, the classical FR BF is not optimal in most cases.

\paragraph{\textbf{3.2 Determination of the specific weight of the priority subsets of the representation forms of BF}}
The specific weight of various SPs in L(n), in particular for n=4, is shown in Fig.2, Fig.\ref{fig3} where that the total capacity of the "classic" subset is seen to be 92.78\% for $S_{ad}$ and only 1.9\% for the most important parameter - $S_s$.\\

\begin{figure}[h!]
\centering{\includegraphics[width=110mm]{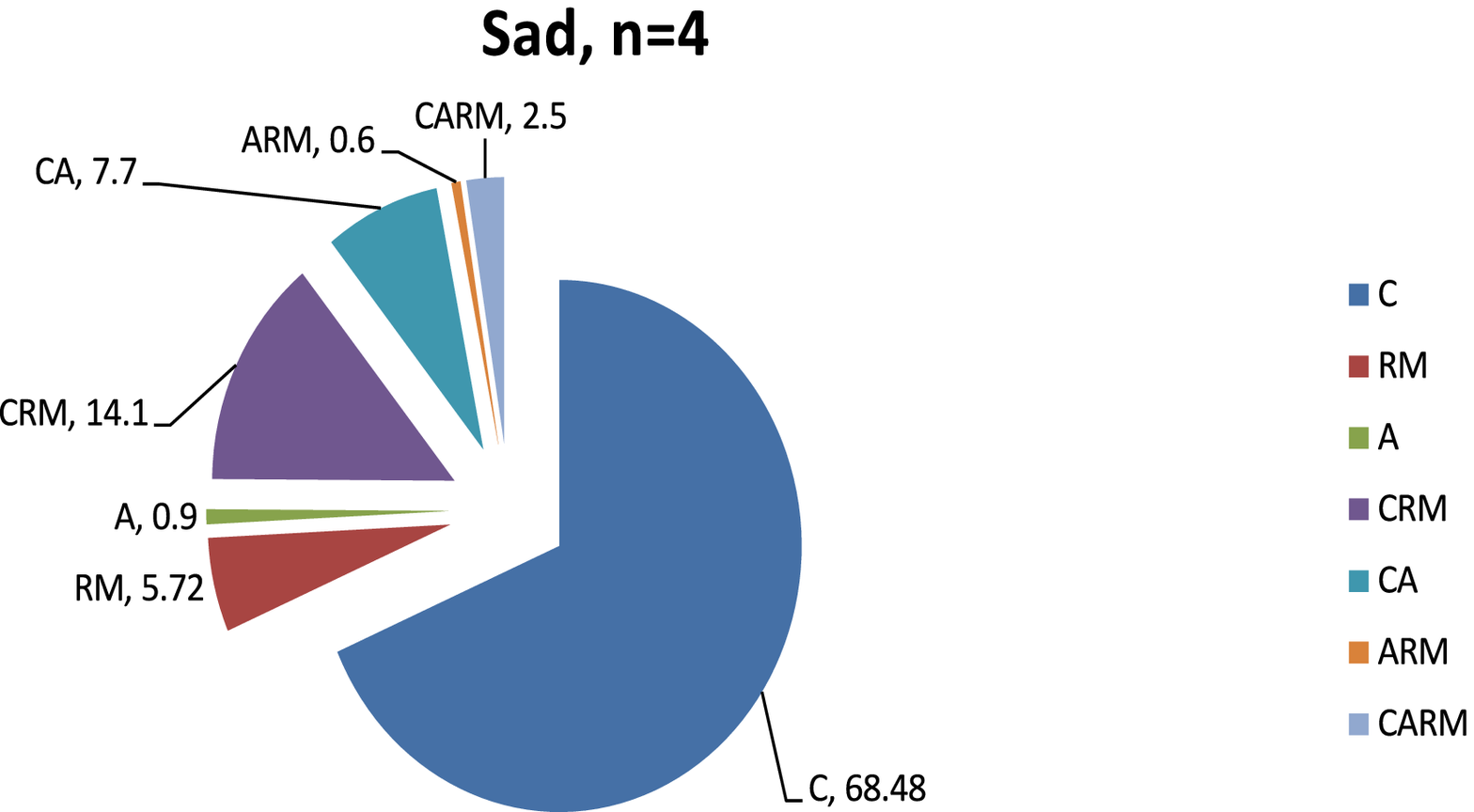}}\\
Figure 2. a)
\end{figure}

\begin{figure}[H]\label{fig2}

\centering{\includegraphics[width=120mm]{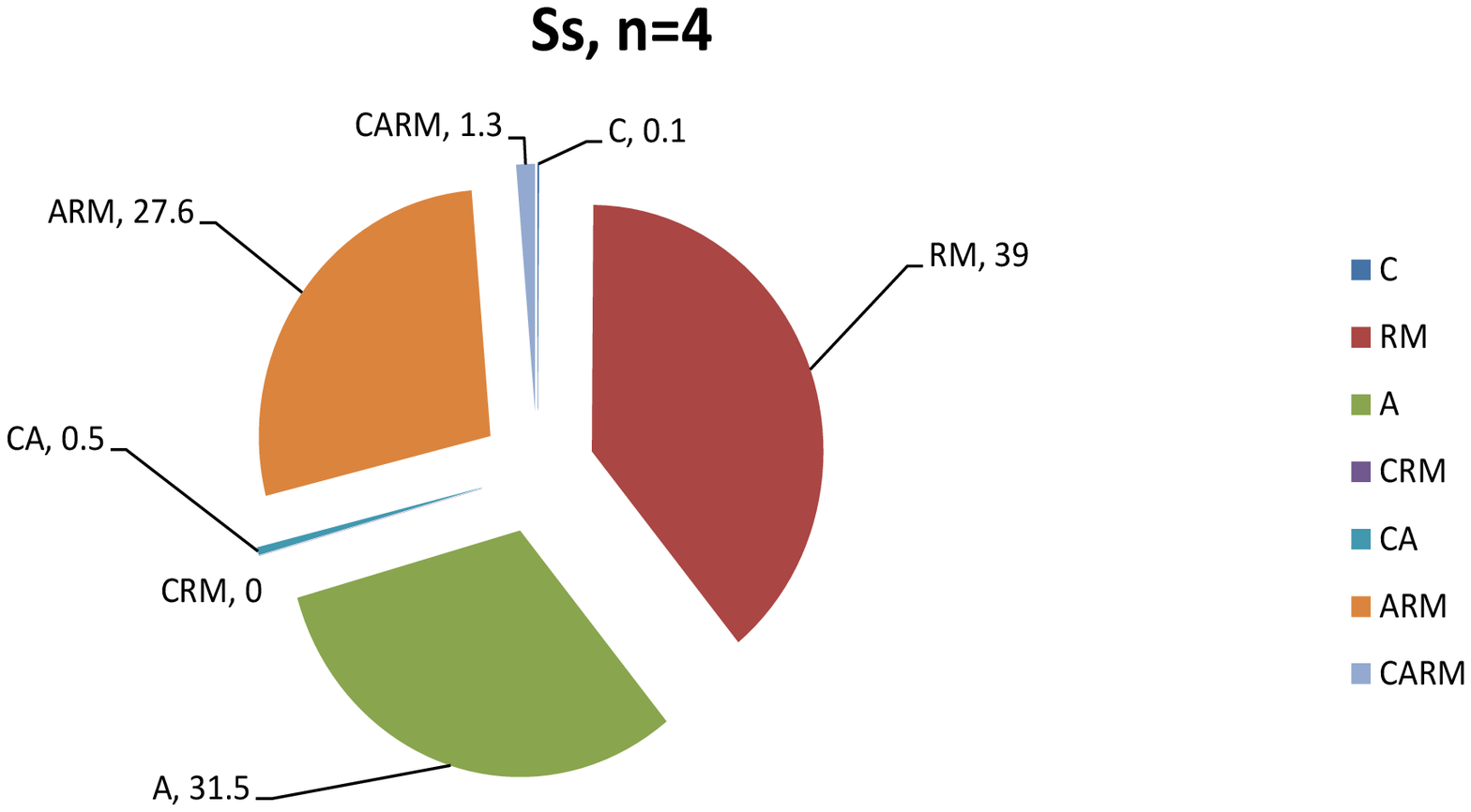}}\\
Figure 2. b)
\caption{\textbf{Specific weight of priority subsets for the parameter L(n): a) Comparison of the $S_{ad}$ coefficient for n = 4; b) Comparison of the coefficient $S_s$ for n = 4.}}
\end{figure}

Here:
-	$S_{ad}$ - the number of summons in the BF record that determines the number of inputs of the summation submatrix of the conjunctions of the PLM2;\\
-	$S_s$ is the averall place of the PLM1, which is determined for both KFR and for AFR and RMFR (n is the number of input arguments of PLM1). This parameter is the most significant, for example, for designing of chips.\\

\begin{figure}[h!]
\centering{\includegraphics[width=120mm]{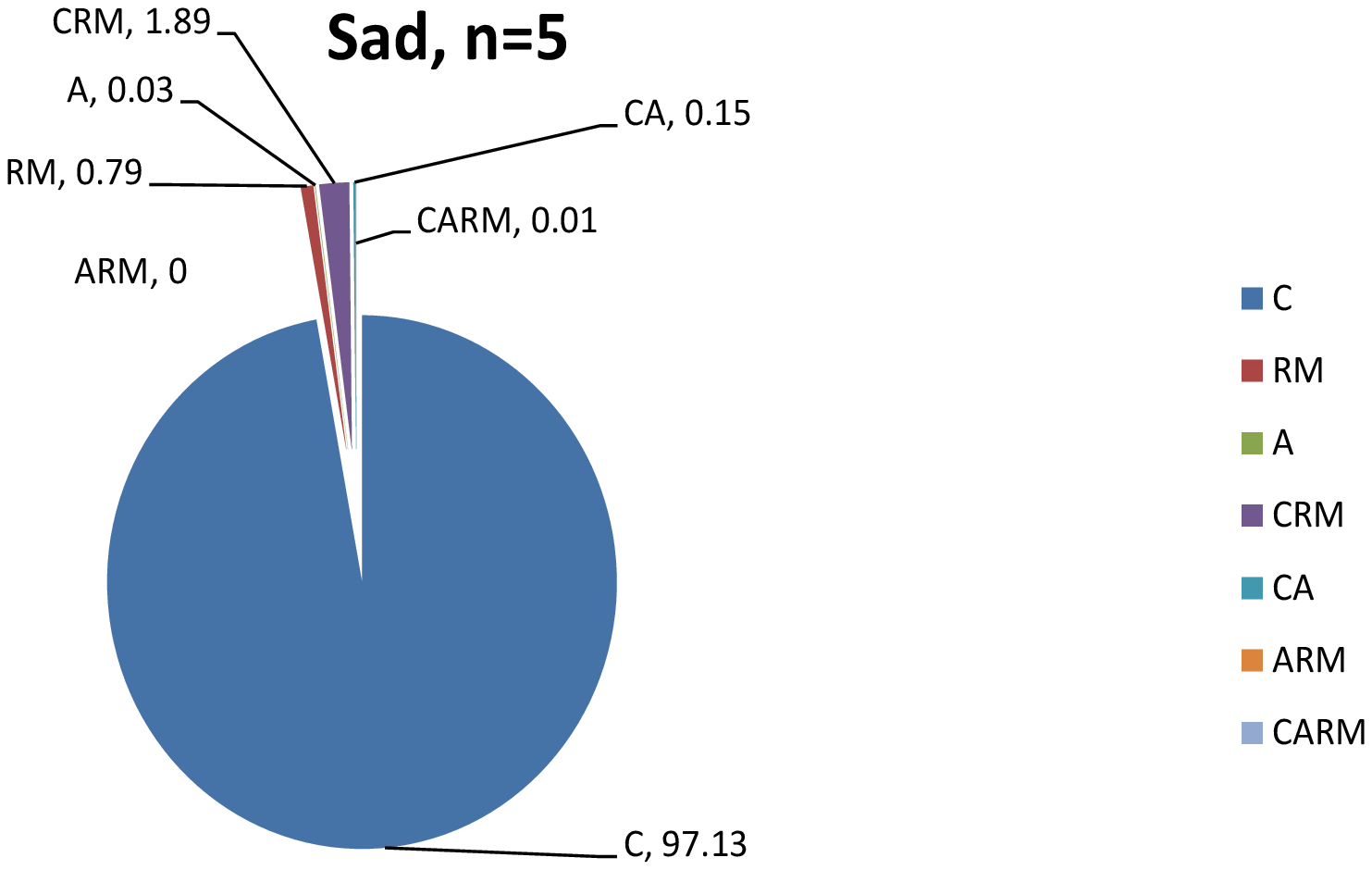}} \\
Figure 3.a)
\end{figure}

\begin{figure}[H]\label{fig3}
\centering{\includegraphics[width=120mm]{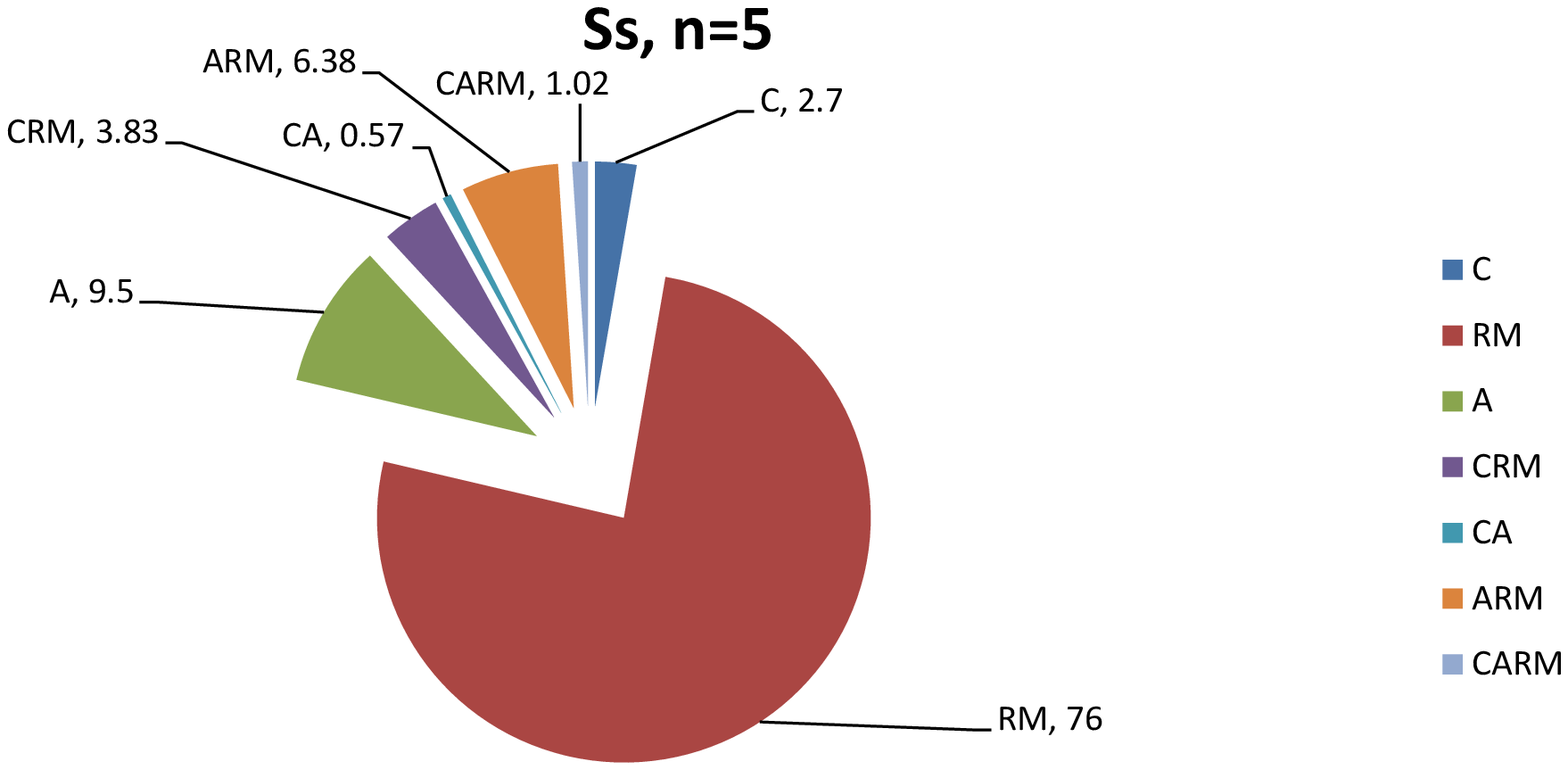}}\\
Figere 3.b)\\
\caption{\textbf{Specific weight of priority subsets for the parameter L(n): a) Comparison of the $S_{ad}$ coefficient for n = 5; b) Comparison of the coefficient $S_s$ for n = 5.}}
\end{figure}

Considering the large number of BFs in L(5) (for n=5), a statistically valid sample of $2^{16}$ BFs is taken, this number guarantees that the inference error will be no more than 5\%.\\
The diagrams also show  that the exclusive use of CFRs, despite the presence of isomorphic AFRs and CFRs, leads to significant losses from the suboptimal use of the FR BF.

\paragraph{\textbf{3.3 Estimation of the statistical average losses from exclusive use of CFR}}
For the following implementation options:\\
- use of only CFR (current situation);\\
- the additional AFR together with the CFR;\\
- the additional RMFR together with the CPR;
- use of all FКs (potential implementation "The concept of ORF").\\
Tables \ref{table3}, \ref{table4}, \ref{table5} present the results of calculating (with formulas \ref{formula_7}, \ref{formula_8}) the estimate for the four implementation variants, respectively, in the complete sets L(3), L(4), L(5).

\begin{table}[H]
\caption{\textbf{Evaluation of benefits for L(3) from the implementation of alternative FRs}\label{table3}}
\begin{center}
\begin{tabular}{p{2.2cm}|p{1cm}|p{1.4cm}|p{1.1cm}|p{1.1cm}|p{1.4cm}|p{1.1cm}}
& \multicolumn{3}{c|}{Parameter Sad} & \multicolumn{3}{c}{Parameter Ss} \\
\cline{2-7}
{Representation form}
& Qad & Absolute benefit & benefit, \%  & Qs & Absolute benefit & benefit, \% \\
\hline
CFR & 590	& -	& 0	& 3540	& -	& 0\\
\hline
CFR +AFR	& 582	& 8	& 1.35	& 2121	& 1419	& 40.1\\
\hline
CFR +RMFR	& 556	& 34	& 5.76	& 2052	& 1488	& 42.03\\
\hline
OFR	& 556	& 34	& 5.76 &	1908 &	1632 &	46.10\\
\end{tabular}
\end{center}
\end{table}

\begin{table}[H]
\caption{\textbf{Evaluation of benefits for L(4) from the implementation of alternative FRs}\label{table4}}
\begin{center}
\begin{tabular}{p{2.2cm}|p{1cm}|p{1.4cm}|p{1.1cm}|p{1.1cm}|p{1.4cm}|p{1.1cm}}
& \multicolumn{3}{c|}{Parameter Sad} & \multicolumn{3}{c}{Parameter Ss} \\
\cline{2-7}
{Representation form}
& Qad & Absolute benefit & benefit, \%  & Qs & Absolute benefit & benefit, \% \\
\hline
CFR & 270897	& -	& 0	& 2167176	& -	& 0\\
\hline
CFR +AFR	& 269633 &	1120	& 0.41	& 1494060	& 673116	& 45.05\\
\hline
CFR +RMFR	& 266113 &	4211 &	1.58 &	1439512	& 727664 &	50.55\\
\hline
OFR	& 265521	& 4695 &	1.73	& 1331348	& 835828	& 62.78\\
\end{tabular}
\end{center}
\end{table}

\begin{table}[H]
\caption{\textbf{Evaluation of benefits for L(5) from the implementation of alternative FRs}\label{table5}}
\begin{center}
\begin{tabular}{p{2.18cm}|p{1cm}|p{1.3cm}|p{1.1cm}|p{1.3cm}|p{1.2cm}|p{1.1cm}}
& \multicolumn{3}{c|}{Parameter Sad} & \multicolumn{3}{c}{Parameter Ss} \\
\cline{2-7}
{Representation form} 
& Qad & Absolute benefit & benefit, \%  & Qs & Absolute benefit & benefit, \% \\
\hline
CFR & 491261 &	-	& 0	& 4912610	& -	& 0\\
\hline
CFR +AFR	& 491236 &	25	& 0.005	& 4528740	& 383870	& 7.81\\
\hline
CFR +RMFR	& 490595 &	666	& 0.135	& 3771185	& 1141425	& 23.23\\
\hline
OFR	& 490570 &	691	& 0.140	& 3716360 &	1196250 &	24.35\\
\end{tabular}
\end{center}
\end{table}

\section{SUMMARY AND DISCUSSION}
From Table.\ref{table3} it can be seen that the parallel use of alternative forms of AFR, RMFR, and also their combination - OFR, allows to slightly decrease the value of the integral indicator $Q_{ad}$ for the set $n=3$ (here this value does not exceed 5.76\%), and for $Q_s$ it is possible to reduce it almost by half in comparison to the "benchmark" CFR - (46.1\%). These results indicate the unjustifiability of only using CFR for the representation of the BF, especially if one considers that it is more expedient for the BF implementation to consider the criterion of decreasing the area of the PLA, and therefore the corresponding index $Q_s$ in this case. In the table we can also see  that RMFR is more economical than AFR.\\
Table \ref{table4} as in Table \ref{table3} for the set L(3) shows continuous improvement of the integral indicators of the structural complexity of implementing BF by the use of alternative forms of representation. At the same time, we should note a slight decrease in the dynamics of the growth in efficiency from the use of alternative forms in the case of the integral indicator $Q_{ad}$ (1.73\%) and the growth in the $Q_s$ saving (62.7\%). From this table we can also see that the RMFR is the closest in terms of indices to the OFR in comparison to AFR. Along with this fact, the big differentiation of the values of the integral indices $Q_{ad}$ and $Q_s$ for the above variants of using the forms of BF representations should also be noted. This fact also supports the use of alternative forms of representation of BF. \\ 
Table. \ref{table5}, as the two preceding tables, confirms the fact that the exclusive use of CFR leads to significant losses of the PLA space associated with certain parameters of the structural complexity of the implementation of the BF. On the other hand, these results show a slight decrease in the effectiveness of BF representation in alternative forms ($Q_{ad}$ - 0.14\% and $Q_s$ - 24.35\%).\\
Thus, it can be concluded that the exclusive use of CFR leads to technically unjustified losses of the chip area, and these losses are quite palpable in absolute and relative sizes, especially in terms of the area of implementation of the BF. Also, the question of further investigation of the FRBF indices with increasing n, remains open. However, this has certain difficulties, since the number of functions for analysis, with n=6, is $2^{64}$ and requires a considerable computational performance. With a further increase in n, an analysis will be possible if statistical methods are applied, however, this will result in certain error in the calculations, which must also be taken into account.

\section*{Acknowledgement}
I thank Dr. Yurii Kochkarev who was the mastermind and
the author of the main idea for this paper.

\end{document}